# Change rates and prevalence of a dichotomous variable: simulations and applications

Ralph Brinks

Institute for Biometry and Epidemiology, German Diabetes Center, Auf'm Hennekamp 65,

40225 Duesseldorf, Germany

Email address: ralph.brinks@ddz.uni-duesseldorf.de

# Abstract

**Background**

A common modelling approach in public health and epidemiology divides the population under study into compartments containing persons that share the same status. Here we consider a three-state model with the compartments: *A*, *B* and *Dead*. States *A* and *B* may be the states of any dichotomous variable, for example, *Healthy* and *Ill*, respectively. The transitions between the states are described by change rates (or synonymously: densities), which depend on calendar time and on age. So far, a rigorous mathematical calculation of the prevalence of property *B* has been difficult, which has limited the use of the model in epidemiology and public health.

**Methods**

We develop an equation that simplifies the use of the three-state model. To demonstrate the broad applicability and the validity of the equation, it is applied to simulated data and real world data from different health-related topics.

**Results**

The three-state model is governed by a partial differential equation (PDE) that links the prevalence with the change rates between the states. The validity of the PDE has been shown in two simulation studies, one about a hypothetical chronic disease and one about dementia. In two further applications, the equation may provide insights into smoking behaviour of males in Germany and the knowledge about the ovulatory cycle in Egyptian women.

**Conclusions**

We have found a simple equation that links the prevalence of a dichotomous variable with the transmission rates in the three-state model. The equation has a broad applicability in epidemiology and public health. Examples are the estimation of incidence rates from cross-sectional surveys, the prediction of the future prevalence of chronic diseases, and planning of interventions against risky behaviour (e.g., smoking).

# Introduction

Most modelling approaches in public health and epidemiology divide the population under consideration into a number of compartments that contain individuals who are identical in terms of their status in question. Famous examples like the SIR-model come from infectious disease epidemiology [1]. The transitions from one compartment to another are described by rates that may depend on calendar time $t$, and in case of age-structured models also on the age $a$ [2]. The model we are dealing with is shown in Figure 1. It consists of the states $A$, $B$ and an additional state *Dead*.

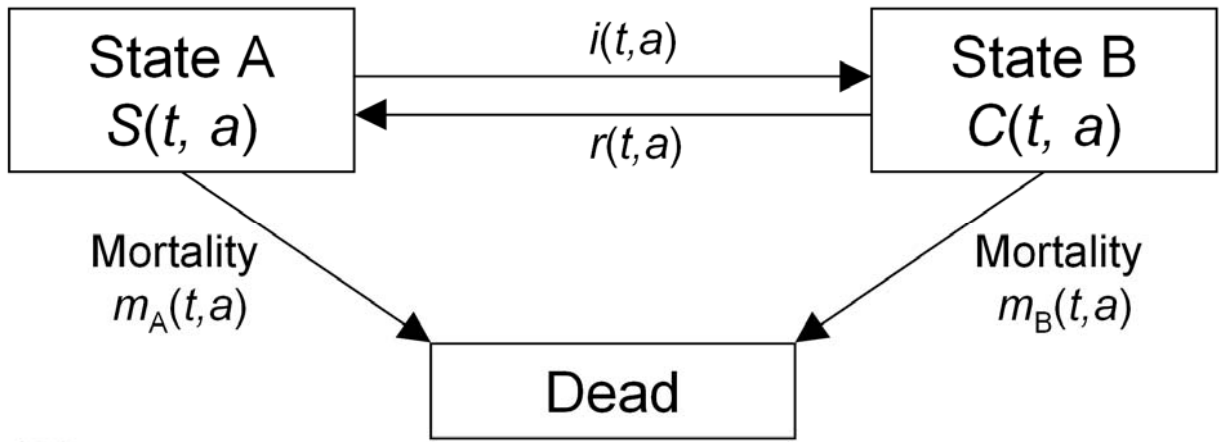

Figure 1: Three-state model. Living individuals of the population aged $a$ at calendar time $t$ are either in state $A$ or state $B$. The respective numbers are $S(t, a)$ and $C(t, a)$. Individuals may change states according to the transition rates.

The model is similar to the illness-death model in epidemiology [3,4], where the states $A$ and $B$ correspond to the states *Healthy* and *Ill* with respect to the disease under consideration. The attributes *Healthy* and *Ill* are a pair of mutually exclusive opposites and describe a dichotomous variable: At a certain point in time, each living person is either healthy or ill (with respect to a specific disease). For now, one may identify states $A$ and $B$ with the *Healthy* and *Ill* state, respectively. Later we will use other dichotomous variables.

The number of individuals aged $a$ at time $t$ in state $A$ and $B$ are denoted with $S(t, a)$ and $C(t, a)$. An important measure in the three-state model is the fraction of (living) individuals aged $a$ at time *t* who are in state *B*. Therefore, define the age-specific prevalence *p* of state *B* as $p(t, a) = C(t, a)/\{S(t, a) + C(t, a)\}$. Of course, this definition is only meaningful if the population has living individuals aged *a* at *t*, i.e. $S(t, a) + C(t, a) > 0$. Furthermore, let the transmission rates be denoted as in Figure 1. All rates may depend on *t* and *a*, and possibly on other covariates.

In case we are dealing with an irreversible state *B*, the rate *r* is zero for all times *t* and ages *a*. In this situation, Keiding has shown that the age-specific prevalence *p* can be written as an integral expression [5]:

$$p(t,a) = \frac{\int_0^a i(t-\delta, a-\delta) \cdot \exp\left(-\int_0^\delta \Psi(t-\delta+\tau, a-\delta+\tau) d\tau\right) d\delta}{1 + \int_0^a i(t-\delta, a-\delta) \cdot \exp\left(-\int_0^\delta \Psi(t-\delta+\tau, a-\delta+\tau) d\tau\right) d\delta}, \quad (1)$$

with $\Psi = m_B - m_A - i$.

Hence, if the rates *i*, $m_A$, and $m_B$ on the right-hand side of Equation (1) are known, the formula allows the calculation of the prevalence of a chronic disease. If future trends in the rates are foreseeable, the future prevalence may be predicted. Moreover, the equation allows to study the impact of interventions on the prevalence. Consider, for example, an intervention that gradually lowers the incidence of type 2 diabetes by 51% [6]. With the assumption, that the mortality rates $m_A$ and $m_B$ are known and remain unchanged by the intervention, the effect of the intervention on the prevalence can be calculated [7]. Then, the age-distribution of the population under consideration can be used to calculate the effect of the intervention on the numbers of cases, its effect on lifetime in health, disability-adjusted life years or other important measures of disease burden. If, in addition, the excess costs of the chronic disease are known, Equation (1) allows the prediction of future disease related costs in a population [8].

The indicated applications show the usefulness of Equation (1) in a variety of health related issues, but despite the usefulness, the formula has rarely been used in epidemiology, public health, health economics or health impact assessment. Apart from the unfortunately low awareness for Equation (1) and the associated theory presented in the seminal work [5], we believe that the equation is possibly too complex to be willingly used by the practitioners in the disciplines mentioned above.

Moreover, we see another fundamental drawback of Equation (1): It is impossible to solve the equation for *i*, which is the age-specific incidence of the chronic disease. The idea of solving Equation (1) for *i*, comes from using cross-sectional surveys to estimate incidence rates, where otherwise lengthy follow-up studies would be necessary [9].

The aim of this work is four-fold:

(I) find a simpler and more flexible expression than Equation (1) that links the prevalence with the transition rates,

(II) generalize Equation (1) to the case of non-zero *r*,

(III) make it possible to extract the rate *i*, and

(IV) show the wide applicability of the theory.

## Methods

### Analytical considerations

We start from the model in Figure 1 and assume that the population is closed, i.e., there is no migration. In addition, all the rates are non-negative and are sufficiently smooth. Similar to the approach in [9], we describe the change rates of the numbers *S* and *C* by differential equations:

$$(\partial / \partial t + \partial / \partial a) \, S = - [i + m_A] \cdot S + r \cdot C, \qquad (2a)$$

$$(\partial / \partial t + \partial / \partial a) \, C = i \cdot S - [m_B + r] \cdot C. \qquad (2b)$$

This means, that the temporal changes in *S* and *C* are the balances of the ingoing and outgoing flows from and to the three states in Figure 1. The result is the two-dimensional system of partial differential equations (PDEs) in (2a) and (2b). A related system has been described by Murray and Lopez [10], however without considering the calendar time $t$. Starting from Equations (2a) and (2b), we derive a new one-dimensional PDE, which fulfils the demands (I) - (IV) mentioned in the introduction.

**Simulations**

To test the validity and the consistency of the resulting PDE, we set up two simulation studies. For simplicity, we choose $r = 0$, i.e. we consider the illness-death model. The first simulation is about a hypothetical population with a priori unknown rates $i$, $m_A$ and $m_B$. The relevant events (birth, diagnosis, death) are generated at random. In the second simulation, we choose the rates $i$, $m_A$ and $m_B$ a priori. Then, we simulate a population with the relevant events birth, diagnosis and death. Hereafter, we try to reconstruct the rate $i$ from the simulated population. The reconstructed incidence is compared to the a priori incidence $i$.

*Simulation 1*

For the first study we simulate a hypothetical disease in a population of one million persons. For each of the persons a date of birth between calendar time $t = 0$ and $t = 100$ (years) is chosen from a uniform random distribution. All newborns are followed from birth to death (without loss). During life-course they may contract the disease or not. Following assumptions are made for the first simulation:

1. The 1,000,000 birthdays are uniformly distributed in the interval (0, 100).
2. At calendar time $t = 0$, the life expectancy (LE) of a healthy individual at the time of birth is 55. This number increases affine-linearly to 75 at $t = 100$.
3. For subjects born at $t = 0$ the lifetime risk for developing the irreversible disease is 10%. This number increases affine-linearly to 25% for those born at $t = 100$.
4. If a subject contracts the disease, the LE lowers by a certain number ΔLE, where ΔLE is uniformly distributed in (5, 10).

5. The age of diagnosis for a subject who contracts the disease during lifetime is X years before death, where X is uniformly distributed in (0, 12).

After the simulation we obtain a set of 1,000,000 persons with dates of birth in (0, 100), an eventual age at diagnosis, and an age at death. If the person does not contract the disease during lifetime, we set the corresponding age at diagnosis entry as missing.

To validate the PDE, we mimic a series of cross-sectional surveys in the years $t_{cs} = 95, ..., 135$ to estimate $p(t_{cs}, a)$, the partial derivate is approximated by a difference quotient. The rates $i$, $m_A$, and $m_B$ are estimated by the person-year method. To check the equality stated in Equation (3), we examine the difference between the left and the right-hand side of the PDE.

*Simulation 2*

The second simulation models dementia in the male German population. From prescribed transition rates $i$, $m_A$, and $m_B$, we simulate a hypothetical register of persons with the relevant events: birth, an eventual diagnosis of dementia, and death. Our aim is the reconstruction of the age-specific incidence $i$ from two cross-sectional surveys.

Unfortunately, the mortality $m_B$ of the diseased and the dependency of the incidence $i$ on the calendar time $t$ are not known for Germany, so we need assumptions to set the simulation up. From life-tables of the Federal Statistical Office in Germany, we get the age-specific general mortality of males in Germany for the past 100 years [11]. The age-specific incidence $i$ of dementia in German males is taken from [12]. Currently, there is no clear indication for a calendar time trend in $i$, so we assume that $i$ does not depend on $t$. In the simulation we follow 457,500 males born in the years 1900 to 1960 from birth to death. As in Simulation 1 there is no loss. Age of diagnosis of dementia and death without disease are modelled as competing risks [13] via an inverse transform sampling from the common cumulative distribution function.

In the years 1998 and 2002, we imitate cross-sectional surveys and measure the age-specific prevalence of dementia in the simulated data. This is done to approximate the partial derivative in the year 2000. From the approximation and the knowledge of the mortality rates $m_A$ and $m_B$ in the year 2000, we reconstruct the incidence $i$. The reconstructed incidence is compared to the incidence that has been used to simulate the population.

**Smoking in Germany**

In contrast to the previous examples, here we do not interpret the model in Figure 1 as an illness-death model, but as a model of a dichotomous health-related outcome. The federal health status report in Germany extensively studied the smoking behaviour of the German population by telephone surveys in the period 1997-1999 [14]. Amongst others, the gender-specific smoking habits of birth cohorts starting from 1921 were analysed. We associate the states *A* and *B* of the three-state model with *Being a non-smoker* and *Being a smoker*, respectively. The definition of being a smoker means regular or occasional consumption of any tobacco products. Although the report covers the age-specific prevalence of (regular and occasional) smoking in 1998, the epidemiologic measures reported in [14] are different from the ones used in Figure 1. For example, instead of age-specific smoking initiation rate (corresponds to the rate $i$ in our model), the mean age at which people start to smoke is reported.

In this application, we ask the question if we can model the age-specific prevalence of smoking in the male German population as reported in [14] by choosing appropriate, i.e. plausible, rates $i$, $r$, $m_A$, and $m_B$ in the setting of our three-state model. As in the previous section, we consider the general mortality from the life tables of the Federal Statistical Office [11].

**Data from the Demographic and Health Survey (DHS) in Egypt**

The set of data for the next application stems from a sequence of six cross-sectional surveys from the years 1988, 1992, 1995, 2000, 2005, and 2008. In the framework of the Demographic and Health Survey (DHS), several thousand Egyptian women were interviewed with respect to many social, demographic and health-related aspects [15]. The property we are interested in is self-reported knowledge about the *ovulatory cycle* (OC). In each of the surveys, all women from a random sample of the female Egyptian population aged 15-49 were asked if there were certain times during the female period in which the chance of becoming pregnant was greater than other times [16]. If the answer was positive, the respondent was asked the follow-up question, during which times this was the case: "during her period", "right after the period", "in the middle of the cycle", "just before her period begins", or "other" (to be specified by the respondent). The dichotomous DHS variable *Knowledge about the ovulatory cycle* is the summary of these questions [16]. A respondent's associated outcome variable was set to be false, if the first question was denied or anything else but "in the middle of the cycle" was replied in the second question.

For our model, states *A* and *B* mean *Does not know about the OC* and *Knows about the OC*, respectively. Based on the age- and calendar-time specific prevalence of state *B*, we want to draw conclusions about the rates *i* and *r* in Egypt, i.e. the learning and the forgetting rates. We assume that the mortalities $m_A$ and $m_B$ are equal for all times and ages.

## Results and discussion

### Analytical considerations

Since we are interested in the prevalence $p = C / (C + S)$, we may apply the quotient rule to the compute the partial derivative $(\partial / \partial t + \partial / \partial a) p$, and insert the expressions for $(\partial / \partial t + \partial / \partial a) S$ and $(\partial / \partial t + \partial / \partial a) C$ from Equations (2a) and (2b). After some straightforward calculations, we obtain:

$$(\partial / \partial t + \partial / \partial a) p = (1 - p) \cdot \{ i - p \cdot (m_B - m_A) \} - r \cdot p. \qquad (3)$$

This means that the temporal change of the prevalence $(\partial/\partial t + \partial/\partial a)\, p$ can be expressed by the right-hand side of Equation (3), which is a combination of the prevalence itself and the four transition rates in the state model. Equation (3) can be interpreted from a birth cohort's perspective. Consider a birth cohort aged $a^*$ at calendar time $t^*$. The prevalence of state $B$ in the birth cohort is $p(t^*, a^*)$. For a small $h > 0$, the change of the prevalence $\Delta_h\, p(t^*, a^*) = p(t^* + h, a^* + h) - p(t^*, a^*)$ is approximately $h$ multiplied with the right-hand side of Equation (3) evaluated at $(t^*, a^*)$.

In case of $r = 0$, a lengthy calculation shows that Equation (1) is the unique solution of Equation (3). Hence, Equation (3) generalizes Equation (1) to the case $r \neq 0$. Another advantage of Equation (3) becomes apparent if the general mortality $m = (1-p) \cdot m_A + p \cdot m_B$ and a relative mortality ratio $R(t, a) = m_B(t, a) / m_A(t, a)$ are given instead of $m_A$ and $m_B$. In this case, an equation equivalent to Equation (3) is

$$(\partial/\partial t + \partial/\partial a)\, p = (1-p) \cdot \{ i - m \cdot p \cdot (R-1) / [p \cdot (R-1) + 1]\} - r \cdot p, \qquad (4)$$

whereas an equivalent version of Equation (1) does not exist. Hence, by the PDE formulation of the relation between prevalence and the change rates, we have a higher flexibility than in Equation (1).

Additionally, Equation (3) generalizes the result described in [9] for the case of rates being dependent on the calendar time. For readers not familiar with differential equations we provide a numerical solver for Equations (3) and (4) in the web appendix of this article. The provided solver is written for the free statistical software R (The R Foundation for Statistical Computing) but may be adapted to other software as well.

## Simulations

*Simulation 1*

We calculate the difference $D(t_{cs}, a^*)$ between the left and the right-hand side of Equation (3) for $t_{cs} = 95, ..., 135$ and $a^* = 5, ... , 85$. Therefore, the partial derivative $(\partial / \partial t + \partial / \partial a)\ p$ at the point $(t_{cs}, a^*)$ is approximated by the difference quotient $\Delta p\ (t_{cs}, a^*) = [p(t_{cs} + h, a^* + h) - p(t_{cs} - h, a^* - h)] / (2 \cdot h)$. Table 1 shows some summary statistics of $D$.

Table 1: Summary statistics of the difference *D*. Summary of the difference *D* between the left and the right-hand side of Equation (3) in Simulation 1.

| Minimum | First quartile | Median | Mean | Third quartile | Maximum | Standard deviation |
|---|---|---|---|---|---|---|
| -0.2007 | -0.0008 | 0.0000 | -0.0002 | 0.0004 | 0.1000 | 0.0060 |

The mean of the difference is close to 0, which indicates the right-hand side of Equation (3) is an unbiased estimator for the partial derivative in (3). To explain the relatively high range of more than 0.3 in the difference *D*, we stratify *D* by the age groups.

The upper part of Figure 2 shows the mean of *D* for each of the age groups $a^* = 5, ... , 85$.

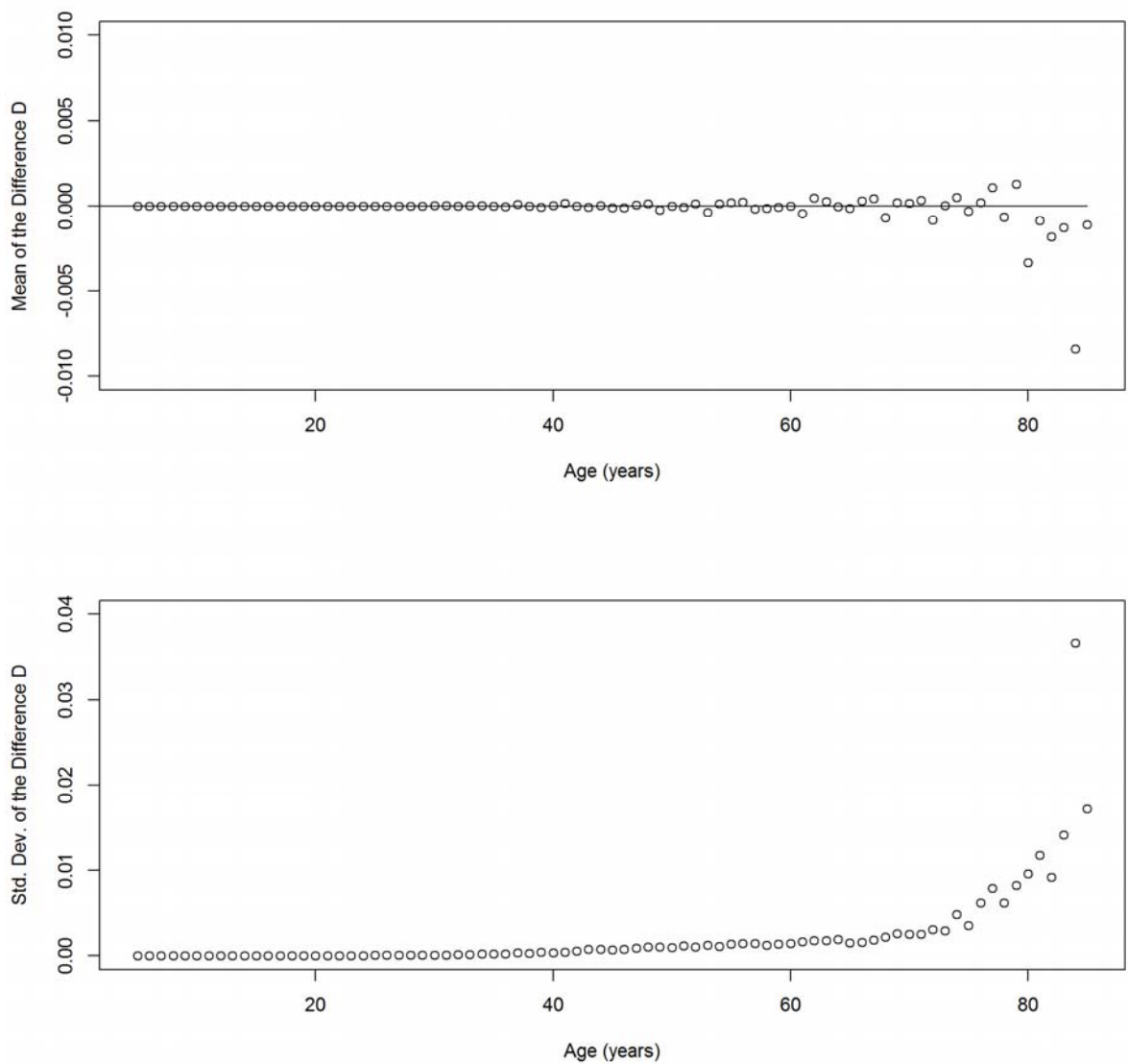

Figure 2: Difference stratified by age. Mean (top) and standard deviation (bottom) of the difference $D$ between the left and the right-hand side of Equation (3) stratified by age $a$.

The values vary around the zero line, again indicating an unbiased estimator. But as the age increases, the variation increases. This increase is also illustrated in the lower part of Figure 2, where the standard deviation of the difference $D$ is depicted versus age $a^*$. Obviously, for higher ages the difference $D$ has a higher variation. The reason for the variation comes from a higher uncertainty of the rate estimation in the higher age groups. For example, in year $t =$ 115, there are 1576 and 88 death cases (with or without the disease) as opposed to 7375.5 and 273.5 person-years at risk in the age groups 70-79 and 80-89, respectively. Calculating the standard error of the person-years method [p. 237, 17] yields a more than 6-fold higher standard error for the (overall) mortality rate in the age-group 80-89 than in the age group 70-79. As a consequence, the difference *D* has a higher variation as the age increases.

*Simulation 2*

For each of the 457,000 persons, we have simulated the date of birth, the age of a possible diagnosis and the age of death. To cross-check these results of the simulation, we compare the age-specific prevalence of dementia in the year 2000 with published values for 2002 [12]. Figure 3 shows the good agreement between the simulated and the reported prevalence.

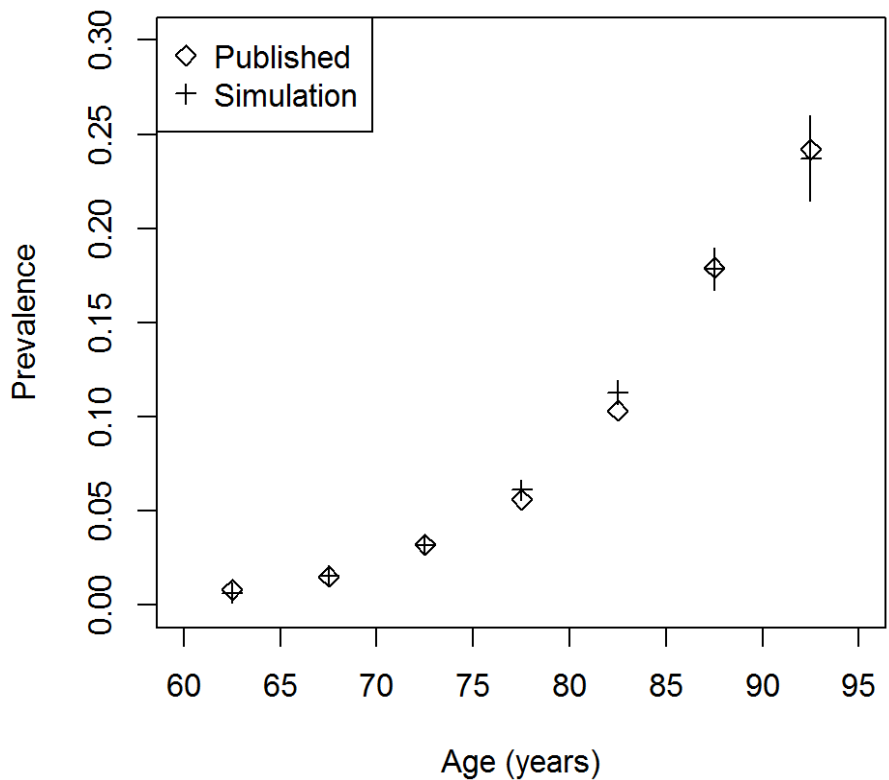

Figure 3: Prevalence of dementia. Age-specific prevalence of dementia in the simulated data (crosses, with 95% confidence bounds) and published values (diamonds, [12]).

Now we take two cross-sectional surveys in the years 1998 and 2002 to get an approximation for $(\partial / \partial t + \partial / \partial a)\, p$ in the year 2000. Together with the mortality rates in 2000, we can solve

Equation (3) for the incidence $i$. Note that we know that $r$ is zero. In Figure 4 the reconstructed incidence is compared to the incidence that has been used for the simulation.

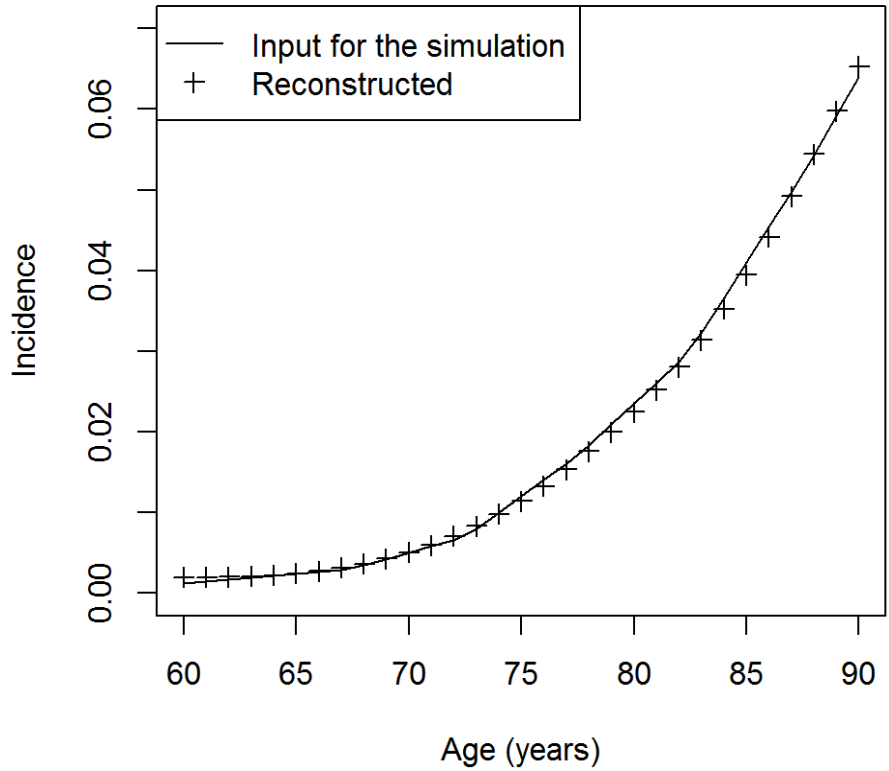

Figure 4: Incidence of dementia. Comparison of the age-specific incidence of dementia: input used to simulate the data (line) and the reconstructed values (crosses).

The reconstructed values follow the age course of the incidence used as input for the simulation very well. Maximum absolute error in the age range 60 - 90 years is $1.53 \cdot 10^{-3}$, and the median of the error is $1.04 \cdot 10^{-4}$. In total, we see a very good agreement between the simulation's input and the reconstructed incidence.

With respect to applications on real data, note that two cross-sectional studies are needed to obtain an approximation for the partial derivative $(\partial / \partial t + \partial / \partial a) \, p$. In case, one of the time dimensions (calendar time or age) is irrelevant, we just need one cross-section [9].

From previous works we know that the quality of incidence reconstruction from cross-sectional data depends on the number of simulated persons in the register [9]. Although the setting therein was slightly different to the one in this work, it is very likely that the accuracy of the reconstructed incidence increases or decreases if the number of simulated persons increases or decreases. An exact study about this would be rather technical and is beyond the scope of this article.

**Smoking in Germany**

Although there are two studies about the mortality of smokers in Germany [18-19], we do not find data stratified by age or even by calendar time. Therefore, we make assumptions for the rates in the three-state model. Since the smoking attributable major death causes (such as neoplasms, cardiovascular and respiratory complications [19]) start after several years of exposure, we assume mortality rates $m_A$ being equal to $m_B$ for ages below 40. At the age of 60 years we assume a relative hazard ($m_B / m_A$) of 2.5, which decreases to 1.5 at age 80. Intermediate values are obtained by affine-linear interpolation. Based on data from three cohort studies in the US reporting values of 1.8, 2.4 and 3.0 in men aged 55-85 [20], we think this is appropriate.

Figure 5 shows the prevalence of smoking in the male German population in 1998. The dark and light grey columns represent the data from the federal health report [14] and from the PDE (3), respectively.

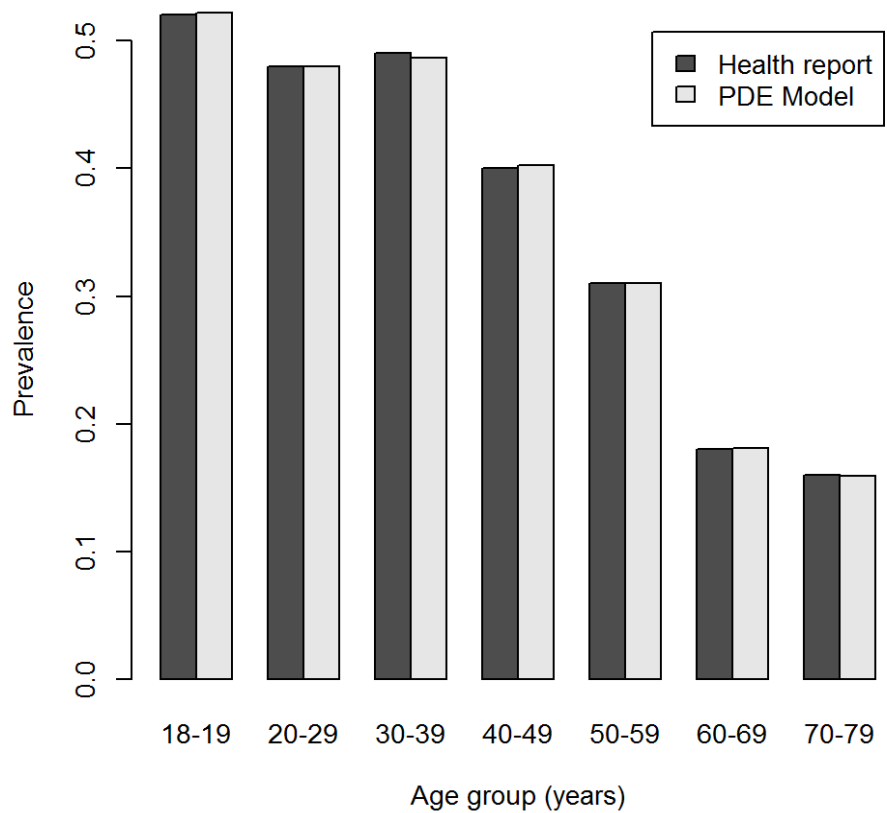

Figure 5: Prevalence of smoking in German men. Age-specific prevalence of occasional or regular smoking in the male German population in 1998. The dark and light grey columns indicate the surveyed and the modelled prevalence, respectively.

Given the general mortality in Germany and the relative mortality hazard, the rates $i$ and $r$ have been chosen such that the data agree quite well. The associated rates are plotted in Figure 6.

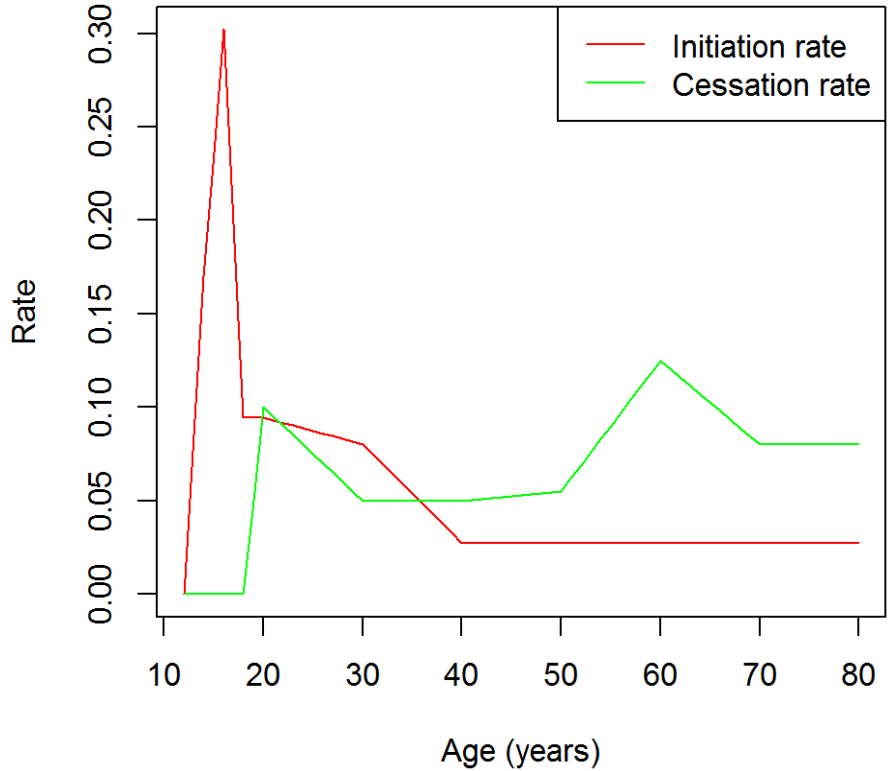

Figure 6: Smoking initiation and cessation rates. Modelled smoking initiation (red) and cessation (green) rates of the male German population in the year 1920. Note that this is only one possibility to model the prevalence of smoking shown in Figure 5.

The rate $i$ of smoking initiation is steeply increasing from ages 12 to 16 and then decreasing as the age progresses. To model the calendar time trend observed in the initiation rate [14], i.e., the dependency of $i$ on $t$, we assume that in each year the rate $i$ decreases by 0.5%.

The cessation rate $r$ is zero until 18 years of age, has a local maximum at the age of 20 and peaks at age of 60. We considered the first peak in $r$, because we think that a lot of subjects try to quit smoking after the teenage time (and maybe restart later in their life). The second peak in the cessation rate at the age of 60 is due to health problems and subsequent medical advice to quit smoking.

Note that we do not state that these rates are real, we are just interested in showing how they could be and that Equation (3) is possible to generate the age-specific prevalence of smoking as in Figure 5 by reasonable assumptions. Other choices of $i$ and $r$ may lead to a similar age course of the prevalence.

**Egyptian DHS**

Figure 7 shows the prevalence of knowing the period of highest fertility during the menstruation cycle (state *B* in our model).

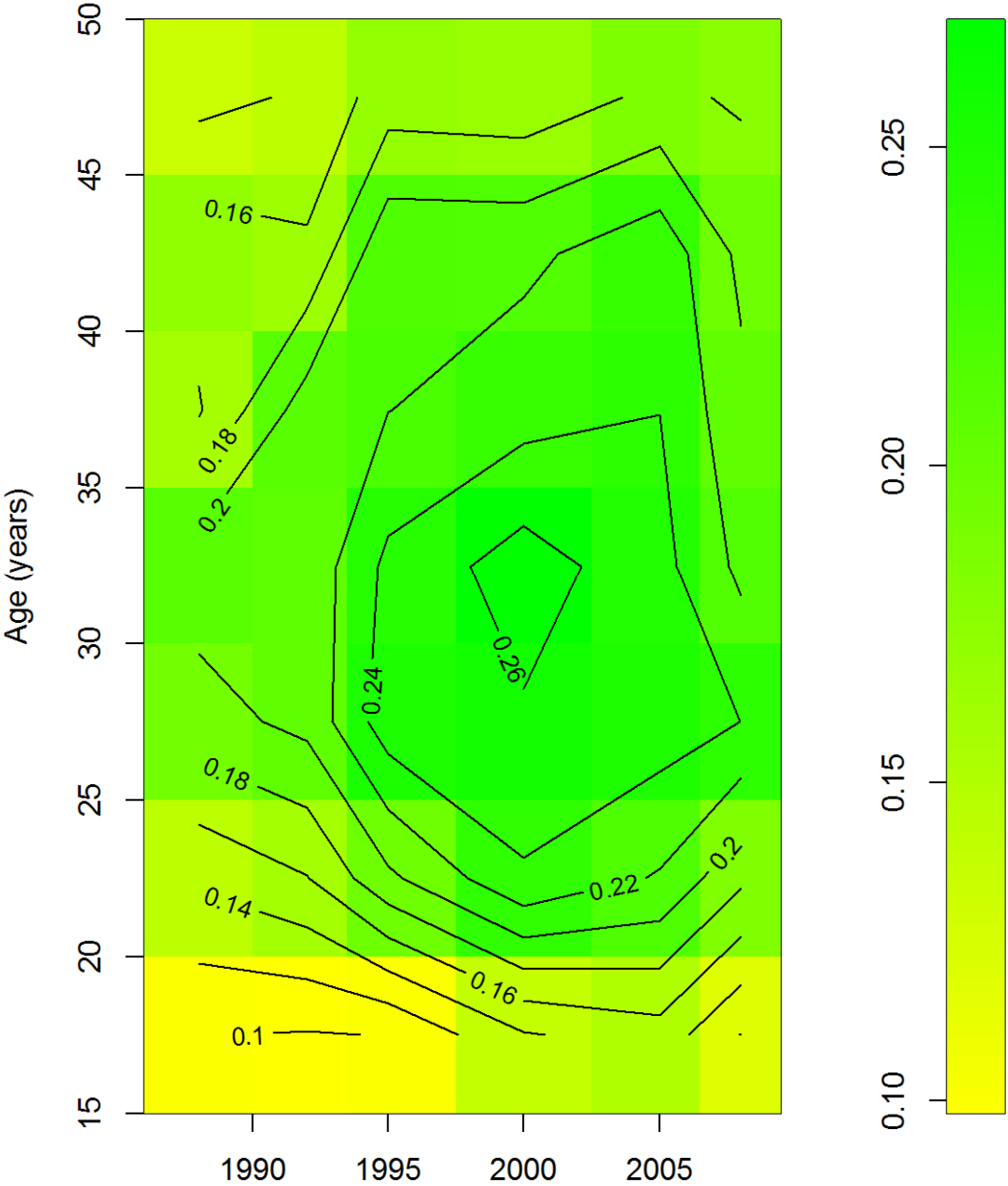

Figure 7: Prevalence of knowledge about the OC. Prevalence of knowing the period of the highest fertility during the ovulatory cycle. Data from six cross-sectional surveys during 1988-2008 in Egyptian women aged 15-49.

From the contour lines it becomes obvious that the prevalence peaks at some age and decreases afterwards. Take, for example, the birth cohort of women aged 20 years in 1988, i.e., all those born in 1968. The prevalence of state *B* is about 12% in that group in 1988. In the year 2000, the persons in the cohort are 32 years of age, and the prevalence is peaking at more than 26%. Some time later, the prevalence in this cohort decreases again. The age course of the prevalence of knowing the OC in this birth cohort is illustrated in Figure 8.

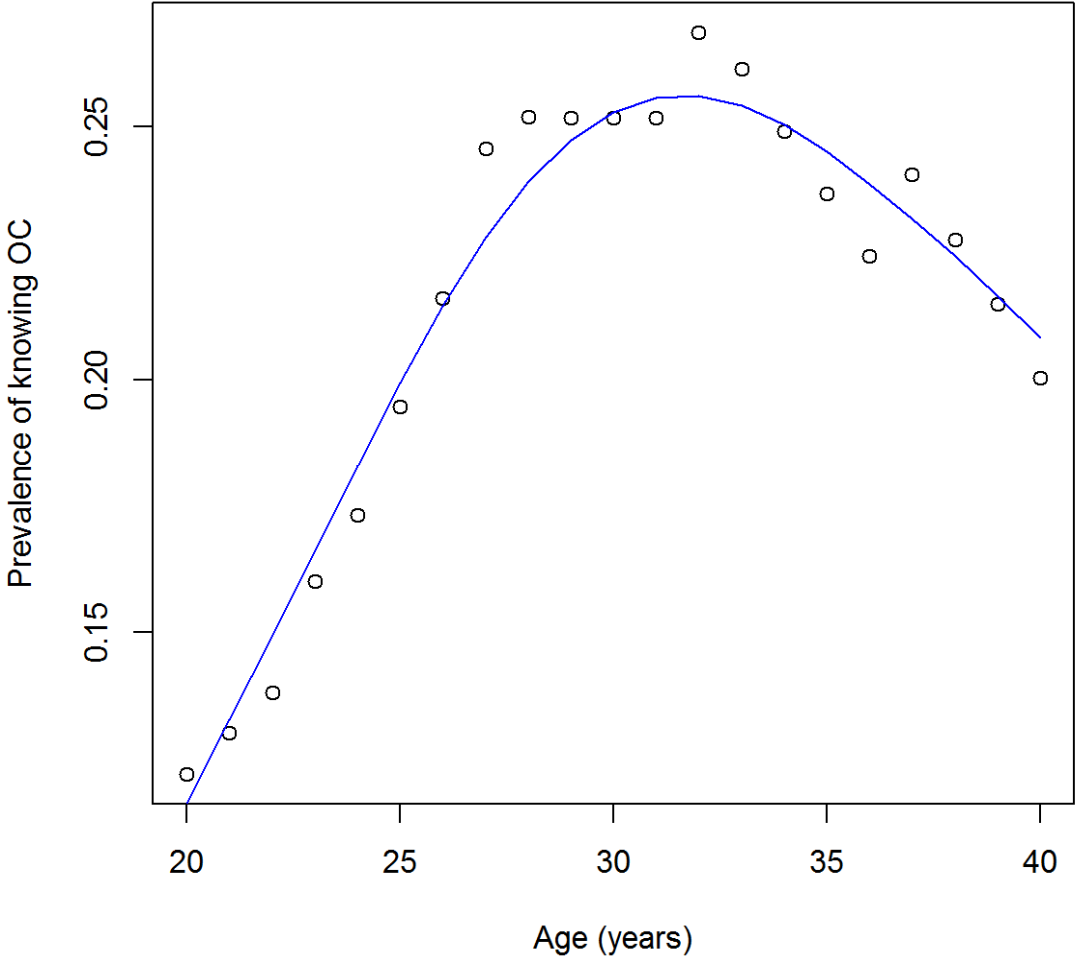

Figure 8: Prevalence in a birth cohort. Age course of the prevalence of knowing about the OC in the birth cohort of Egyptian women aged 20 in 1988.

The decrease after the age of 32 implies that those women who once knew about the nature of the OC, start to forget about it. All birth cohorts in the study have this pattern of learning until a certain age (mostly around 30-40) and forgetting afterwards. This justifies the following approach: If the partial derivative $(\partial / \partial t + \partial / \partial a)\ p$ is positive, i.e. the prevalence in the birth cohort increases, we solely attribute the increase to $i$; if the partial derivative is negative (the prevalence in the birth cohort decreases), we attribute the decrease solely to $r$. Tables 2 and 3 show the results about the learning and forgetting rates $i$ and $r$.

Table 2: Learning about the OC. Rate of learning about the period of highest fertility during the ovulatory cycle (in units 1000 per person-year) in Egyptian women stratified by year and age.

|      | Age (in years) | | | | | | |
|------|----|----|----|----|----|----|----|
| Year | 15 | 20 | 25 | 30 | 35 | 40 | 45 |
| 1988 | 15.2 | 18.6 | 12.9 | 4.1 | 3.3 | 1.1 | 0.0 |
| 1993 | 22.0 | 21.4 | 19.9 | 10.8 | 3.8 | 5.0 | 0.0 |
| 1998 | 26.0 | 26.9 | 16.8 | 5.5 | 0.0 | 0.0 | 0.0 |
| 2003 | 23.2 | 17.6 | 5.6 | 0.0 | 0.0 | 0.0 | 0.0 |
| 2008 | - | 2.5 | 3.3 | 0.0 | 0.0 | 0.0 | 0.0 |

Table 3: Forgetting about the OC. Rates of forgetting about the period of highest fertility during the ovulatory cycle (in units 1000 per person-year) in Egyptian women stratified by year and age.

|      | Age (in years) | | | | | | |
|------|----|----|----|----|----|----|----|
| Year | 15 | 20 | 25 | 30 | 35 | 40 | 45 |
| 1988 | 0.0 | 0.0 | 0.0 | 0.0 | 0.0 | 0.0 | 35.5 |
| 1993 | 0.0 | 0.0 | 0.0 | 0.0 | 0.0 | 0.0 | 24.8 |
| 1998 | 0.0 | 0.0 | 0.0 | 0.0 | 0.2 | 6.8 | 35.0 |
| 2003 | 0.0 | 0.0 | 0.0 | 11.4 | 23.9 | 21.7 | 41.0 |
| 2008 | - | 0.0 | 0.0 | 13.5 | 39.5 | 39.4 | 42.8 |

We can see a two general trends in the tables. First, from top left to the bottom right - this is the direction of progressing time - the learning rate decreases and the forgetting rate increases. Second, the learning rates are lower than the rates of unlearning. It seems that knowledge about the ovulatory cycle is gained slowly until a certain age, about 30 to 40 years, then it is disremembered very quickly. As we are just interested in showing applicability of the method, we do want to not discuss these results more in depth.

We conclude this section with some remarks about the assumptions used to analyse the data by Equation (3). First, we look at the approach of strictly attributing a positive and negative partial derivative to learning and forgetting. Of course, this assumption seems to be unrealistic on the individual level of the members of a birth cohort. Some members of the cohort might still plan to become pregnant and seek for information about time periods of more probable conception, i.e. they learn, while other members of the cohort have already finished their family planning, have decided against (further) child bearing and start to unlearn. Especially during the peak of the prevalence, both processes, learning and unlearning, take place synchronously within the birth cohort. By the strict attributing changes of the prevalence to either *i* or *r*, we are focussing on the "average" trend of the birth cohort, not on the individual level. However, from a public health perspective we typically are interested on the "average" behaviour of a cohort, for example, in planning an education campaign about contraception. Thus, we think focussing on the cohort average is useful and reasonable.

The second assumption that has been made is that the mortality rates $m_A$ and $m_B$ are equal. One might argue that the mortality $m_A$ of the "uninformed" women is higher than $m_B$, for example by hypothesising that not knowing about the OC is associated with not knowing about the human body, about its physiological processes and its needs at all. Thus, one might conclude that the uninformed women have an unhealthier life style and thus a higher mortality. Not knowing about the OC may also be associated with living in poverty or living in rural areas. For those persons, the medical care is likely to be worse than the average and thus the mortality $m_A$ might be higher than $m_B$, too.

On the other hand, it might be possible that the informed women have a higher mortality than the uninformed. Knowing about the OC may be positively correlated with a riskier behaviour with respect to contacts with males. Someone who knows that it is less likely to get pregnant at certain periods during the cycle, might be willing to use condoms less frequently and might be more exposed to sexually transmitted diseases. In that respect one might think of HIV, but Egypt has a relatively low prevalence of HIV (below 0.2 per mille in 2009), however, with a time trend pointing upwards [21].

Whatever plausible explanations may be found for the hypotheses that $m_A$ is different from $m_B$, empirical data is unlikely to ever exist. For this reason, we assume that $m_A$ is equal to $m_B$.

## Conclusions

This article is about a new partial differential equation (PDE) that models the prevalence of a dichotomous variable in a population as a function of calendar time and age. The most frequent use of the model is the illness-death model where the dichotomous variable means healthy or ill with respect to a specific disease. The model comprises the case of differential mortality depending on the current disease status. As an application of the PDE in the context of illness-death models, we have examined two simulation studies: One about a hypothetical disease and one about dementia in the male population of Germany.

Although the illness-death model may be an important application, the PDE is not restricted to disease states and may be used in any other dichotomous variable, property, factor or outcome. Two outcomes studied in this article are smoking behaviour of German males and knowledge about the ovulatory cycle in Egyptian women. Other interesting dichotomous variables that can be modelled by the PDE are easily imaginable: addiction to alcohol or drugs, risky behaviour in traffic, at work or in sports (e.g., the non-use of seatbelts, helmets, protectors etc) or knowledge about harmful substances in food or in the environment. In these regards, the PDE may be used to model and predict the impact of changes in the transition rates between the states on the prevalence of the property. To give an example, we consider the situation that the health authorities have to choose between two different programmes in smoking prevention. Assume that due to budget restrictions the authorities had to decide for one of the programmes. The first programme lowers the initiation rate by a media campaign and the second programme rises the cessation rate by putting nicotine substitutes onto the formulary. Assumed we know the age-specific rates and the effects of both programmes exactly, and the outcome of interest is the number of smokers in, say, five years. Then, we can use the PDE to predict impact of the programs on the age-specific prevalence in five years. By multiplying the predicted prevalences with the projected age distribution of the population, we can compare the programmes.

Another application of the PDE may be seen in health registers or large databases of health insurances. Some countries have registers for all persons with a diagnosis of a chronic disease, for example diabetes in Denmark [22]. The incidence and mortality rates of the illness-death model are easily accessible in these registers. A question may be how the prevalence of the disease in the future will develop, if trends or changes in the rates are extrapolated. In data from health insurances, a variety of aspects may be examined. To give an example, one might be interested in the age-specific prevalence of insured persons being in a hospital at a specific point in time. The PDE allows a straight-forward calculation of this prevalence from the rates in the three-state model.

With a view to the broad applicability of the three-state model, we think that Equation (3) will help to model and predict a variety of epidemiological and public health outcomes.